# Evolving Academia/Industry Relations in Computing Research

Shwetak Patel (Univ. Washington), Jennifer Rexford (Princeton Univ.),
Benjamin Zorn (Microsoft), Greg Morrisett (Cornell Univ.)
Industry Working Group, Computing Community Consortium (CCC)
June 2019

## Executive Summary

In 2015, the CCC co-sponsored an industry round table that produced the document "The Future of Computing Research: Industry-Academic Collaborations."[1] Since then, several important trends in computing research have emerged, and this document considers how those trends impact the interaction between academia and industry in computing fields. We reach the following conclusions:

- In certain computing disciplines, such as currently artificial intelligence, we observe **significant increases in the level of interaction between professors and companies**, which take the form of extended joint appointments.
- Increasingly, **companies are highly motivated to engage** both professors and graduate students working in specific technical areas because companies view computing research and technical talent as a core aspect of their business success.
- There is also the further potential for principles and values from the academy (e.g., ethics, human-centered approaches, etc.) informing products and R&D roadmaps in new ways through these unique joint arrangements.
- This increasing connection between faculty, students, and companies **has the potential to change (either positively or negatively) numerous things, including**:
    - The academic culture in computing research universities
    - The research topics that faculty and students pursue
    - The ability to solve bigger problems with bigger impact than what academia can do alone
    - The ability of universities to train undergraduate and graduate students
    - How companies and universities cooperate, share, and interact
- This report is the first step in engaging the broader computing research community, raising awareness of the opportunities, complexities and challenges of this trend but further work is required. **We recommend follow-up** to measure the degree and impact of this trend and to establish best practices that are shared widely among computing research institutions.

---

[1] https://cra.org/ccc/industry/
[2] https://cra.org/data/generation-cs/
[3] "The Hard Part of Computer Science? Getting into Class", New York Times, Jan. 24, 2019, https://www.nytimes.com/2019/01/24/technology/computer-science-courses-college.html

**Introduction**

Several notable trends have changed the landscape of computer science education and research in recent years. First, undergraduate computer science enrollment has increased dramatically.[2,3] Second, information technology has become a product differentiator in almost every industry, leading to the broad sense that "All companies are technology companies now."[4] Advances in AI and deep learning, where research is expanding at an exponential rate, have led to large increases in AI investment.[5] For many applications of computer technology, including the Internet of Things (IoT), healthcare, and autonomous vehicles (AVs), the demands on software technology in terms of creative new solutions, correctness, safety, security, privacy, etc. have significantly increased. Faculty are now being sought after to not just collaborate in research, but to drive and lead innovative new efforts within industry. In the words of Karl Iagnemma (President, Aptiv Autonomous Mobility), speaking about companies investing in technology, "Previously [computer science] research was optional, now it is not." This is certainly an exciting time in computing research as it relates to industry interactions and collaborations.

Given these dramatic changes, it is natural to wonder how these trends impact the relationship between universities doing computing research and companies using it. The CCC published a report in 2015 that captures many important aspects of how academia and industry have typically interacted in the past. The goal of this document is to discuss how things have changed since 2015 and consider ways to steer those changes to have the greatest benefit for both academia and industry. Over the last year we have conducted numerous interviews with senior academics, industry, and government officials trying to answer four questions:
- Has the relationship between academia and industry in computing research changed in recent years?
- If it has changed, what are the potential impacts of those changes?
- Given these impacts, what can be done to make them the most positive and constructive for both academia and industry?
- What are some guidelines we can offer to academia and industry, specifically to increase innovation and impact, protect the interests of the students, and to preserve the culture of freedom and inquiry in academia?

---

[2] https://cra.org/data/generation-cs/
[3] "The Hard Part of Computer Science? Getting into Class", New York Times, Jan. 24, 2019, https://www.nytimes.com/2019/01/24/technology/computer-science-courses-college.html
[4] https://blogs.thomsonreuters.com/answerson/all-companies-are-technology-companies-now/
[5] "10 Charts That Will Change Your Perspective on Artificial Intelligence's Growth", Forbes, Jan. 12, 2018. https://www.forbes.com/sites/louiscolumbus/2018/01/12/10-charts-that-will-change-your-perspective-on-artificial-intelligences-growth/#59968dfa4758

Our findings are based on conversations and interviews from various stakeholders in academia and industry. Although we provide some suggestions and guidelines, we do not have detailed objective measures of the significance of these trends, and we hope this document encourages more detailed data collection and conversation on the topic. To focus our conversations and as a case study, we began by looking at a specific subarea of computing research that has been in the public eye--transportation innovation and autonomous vehicles--but quickly expanding beyond transportation. While the trends we discuss do not apply equally across all computer science disciplines, we see similar trends in computing research related to IoT, health, and AI. We also discuss other benefits joint appointments can have industry and academia as it relates to new perspectives to help drive innovation and research.

**The Trend: Increases in Faculty Joint Appointments**

Based on our conversations, we conclude that the relationship between (some parts of) academia and (some parts of) industry is evolving rapidly. In particular, we see significant increases in the level of interaction between professors and companies, greater efforts from companies to deeply engage both professors and graduate students, and increased complexities for universities to understand and manage this interaction, including the impact it has on the university culture and education mission. Across the nation, we have seen a significant number of our computing research faculty on extended partial or full-time leave in industry and startups.

Across a number of academic institutions, we were informed of several patterns:
- The **number of faculty holding joint or concurrent appointments has increased**.
- The split in joint appointments has shifted to allow **professors to spend more time at companies**. Where previously the common allocation was 1 day a week consulting (20/80) plus summer, now appointments are often 50/50 or even 80/20 (1 day at the university).
- Even without spending more time (with a 20/80 appointment), professors are strongly aligning with companies and **signing employee contracts** with certain non-disclosure and non-compete requirements.
- The duration of such appointments is frequently not limited, allowing for **indefinite periods where faculty are part time** at the university in some cases.

One reason for this shift is that, due to trends in cloud computing, big data, health, and AI, **many important computing research problems require resources at a scale now unavailable to academics**. The increase in joint appointments reduces the time a faculty member spends at the university and has implications on many aspects of their academic responsibilities.

**Implications of Greater Faculty/Industry Connections**

As with any major shift, the implications of stronger ties between faculty and companies can be positive or negative. Our goal is to explore both positive and negative impacts and try to understand ways to increase the benefits while mitigating the risks. We consider several dimensions of impacts below.

**Department culture and service:** An essential job of any professor is mentoring graduate students and working with colleagues to create a culture of enrichment in their department that enables students to grow, support each other, and develop their creativity and curiosity to enable them to do great research. Creating this culture requires faculty effort in connecting students with each other, providing opportunities to expand their experience via coursework, department seminars, and other activities, and the ongoing effort of attracting and recruiting new students. Companies contribute to department culture in many ways as well, including supporting fundraising, helping make connections across an institution, and across institutions. Nevertheless, having a professor working only 50% time or 20% time at a university has an impact on their ability to maintain such an environment for students that is hard to measure but also hard to ignore. On top of that, it is important to recognize service is critical to the operation of a department which includes student/faculty recruiting, planning, broader university connections, fundraising, etc. There appears to be an explicit need to set expectations on the faculty member's engagement with the department and university while on partial leave or serving a joint appointment.

**Research agenda:** The research agenda of professors and graduate students is subtly shaped by numerous influences including grant support, industrial support, and shifting interests of the professor and student. Collaborating with companies often can enhance a research agenda by injecting important real-world problems, increased resources including computing and data, and access to highly skilled engineers often not available at universities. The challenge around joint appointments where a professor spends a significant fraction of their time working at a company is that the level of expectations the company has for the professor increases and the expectation of the company might be that the professor coax their research agenda (including possibly that of their students) toward the product needs and timeframes of the company.

**Conflict of interest (COI):** While issues related to COI are not new (since historically start-ups have been a common vehicle for CS faculty to transfer their research to practice), the degree of engagement and the affiliation of professors with corporate entities increases the complexities of handling COI. The alignment of a professor's interests with those of a company increases the potential for conflict of interest. This conflict can take many forms including impacting what research a professor does (see above), influencing the research of their students, influencing their ability to evaluate academic work from a competing company, and creating confusion when speaking or publishing in an academic forum as to whether they are representing the company or providing an unbiased academic perspective. In medical journals, such as the Journal of the American Medical Association, the requirements for specifying the financial interests of the authors are much stronger than they are in computing research.[6]

**Changes in How Companies Engage with Graduate Students**
Often graduate students spend summers during their PhD doing three-month internships with companies as a way to expand their research experience, connect with a different set of interesting problems, and learn more about the companies themselves. While internships are valuable to students, they also often shift the student's focus away from their PhD research. The three-month duration allows faculty to balance the student's focus on PhD research against the other benefits internships provide. Moreover, companies have recently incentivized internship by raising their pay relative to research assistant salaries.

With professors having increased connection to companies, it is also the case that the amount and degree of interaction of graduate students and companies has increased. Specifically:
- Graduate students have more extended interactions with companies (beyond three month internships). These include extended contractor positions, longer co-op hiring arrangements, and even joint advising agreements where staff at companies co-advise a PhD student. Often, extensions arise necessarily to allow students to complete research started during internships.
- If a faculty member's research focus is biased toward problems related to a company, graduate students may increasingly have their research aligned with a company's goals.
- Because companies have resources that extend beyond what is available to an academic, a graduate student's degree may increasingly depend on maintaining a relationship with a company.

Because the competition for the talent of PhD students is increasing, companies have strong incentives to engage graduate students early and connect with them throughout their graduate career. Having hired a faculty member on a 50+% time contract, it is a natural outcome that both the faculty member and the company have a significant interest in the professor's students working for the company as well. There are substantial potential benefits to students who pursue extended work agreements with companies (either through contracts or multiple internships), but there are also potential risks.

**Shift in research emphasis**: Companies working closely with professors will have a direct impact on their research agenda. In many cases, this impact can be positive with the company helping a professor understand what the important problems are and bringing new resources that allow bigger problems to be solved. Often companies face important technical challenges that professors are not even aware of. Alternately, the practice of graduate students working on company problems has the potential to shift the research emphasis of the student away from longer-term research and toward short-term results that have immediate benefits to the company. Because companies are driven by competition in the market, graduate students may find the focus of their work more driven by market needs and less by doing the highest impact research. More broadly, high visibility collaborative research at companies can influence the research agenda of an entire field. With the potential for both positive and negative impacts on research emphasis, care must be taken to avoid the negative outcomes by anticipating and avoiding them.

---

[6] Phil Fontanarosa and Howard Bauchner, "Conflict of Interest and Medical Journals", *JAMA*. 2017;317(17):1768-1771. https://jamanetwork.com/journals/jama/fullarticle/2623590

**Shift in research supervision**: One approach that reduces the burden of lack of student supervision created by faculty joint appointments is for companies to supply individuals to co-mentor or entirely mentor the PhD work of students[7]. Such supervision has a number of advantages, including providing the student with a better understanding of the problems companies face, how technology is developed at companies, and what a long-term career in the field is like. Risks associated with this practice include the possibility that the mentoring individual does not have the necessary qualifications to supervise PhD research and whether the individuals are sufficiently aware of potential conflict of interest issues with respect to acting in the best interests of the student. For this approach to be successful, any company mentor has to understand that students are first and foremost students (not employees) and their university obligations prevail over anything else.

**Conflict of interest and power differential:** With a faculty advisor and/or a company staff mentor with a significant economic interest in the success of the company, the potential for conflict of interest, where the student's needs are secondary to the success of the company, arises. This COI challenge is particularly important with faculty and graduate students because faculty have such a large degree of power over the lives of their students. Universities are responsible for protecting students in such situations and ensuring that they receive the best mentoring and guidance possible. Ultimately, the quality of the research done by graduate students impacts the quality and reputation of the university overall. As a result, universities need to maintain a clear understanding of faculty/student interactions and provide best practice COI policies to protect students.

**Reduction in graduate student connection:** Beyond the impact on research choices and directions, the increasing engagement of faculty with companies has other lasting implications for graduate students. Students spending time working at a company may engage less with other students at the university but at the same time they will be exposed to a different group of students (interns from other universities, etc.) and be more directly exposed to the corporate community and culture. Faculty spending less time at the university constrains the amount of interaction graduate students have with their advisors, even if the students are not working on company-related research. Graduate students who do work on company-related research for a professor may receive preferential treatment with respect to amount of contact and mentoring. Anecdotally, we are aware of cases where students failing to work on company-related problems quietly complain about their lack of faculty access or students may choose to work on related problems for more face time with the faculty. Less student contact reduces the ability of the professor to closely follow and direct student research and may also result in reducing the connection between students working for the same professor.

---

[7] Outside the United States, such arrangements are more common. For example, the Industrial PhD Program at the University of Copenhagen.
https://www.science.ku.dk/english/research/phd/studystructure/industrialphd/

**Implications for Undergraduates**

As mentioned earlier, the significant increase in undergraduate CS enrollment places stresses on department teaching resources and results in larger class sizes, increased use of undergraduate teaching assistants, and decreased contact between faculty and undergraduates. Adding a shift in faculty joint appointments on top of that trend reduces the amount of investment a professor may have on curriculum development and undergraduate research. In addition, because graduate students play a critical role in supporting the undergraduate education and research mission, if they are also more deeply engaged with companies they may also contribute less to training undergraduates.

While we have little anecdotal or objective data about the undergraduate implications of the shift in faculty engagement with industry, the following questions are relevant:
- Have interactions with companies provided new opportunities for undergraduates to pursue interesting research at companies or better career choices?
- Does having a significant joint appointment reduce the engagement of professors with undergraduates, including the support of undergraduate research?
- Has undergraduate interest in a research career increased or decreased?
- Has the amount of contact that undergraduates have with faculty decreased over time?
- Has the amount of time graduate students interact with undergraduates decreased?

An ongoing concern about companies more deeply engaging faculty is that the long-term effect will be to reduce the quality of training of future students at all levels, especially if fewer individuals choose to pursue an academic career as a result. Because these effects can happen over time, it is important to observe these trends and take measures to prevent them, starting with the impact on undergraduates. Because companies have a vested interest in maintaining a strong talent pipeline in computing research, they should be especially interested in efforts to measure and enhance the process.

Inspiring and educating undergraduates about what research is and what is involved in getting a PhD is an important part of a computing research university culture. If, for example, undergraduate research positions are fewer and more competitive, it will serve as a disincentive for students to participate. Another direct impact of making undergraduate research opportunities more competitive might be to hinder efforts to broaden participation in computing research for undergraduates, an outcome that should definitely be avoided. Companies have both the incentive and opportunity to enable universities to attract and engage undergraduates in the research process.

**Trends in Industry Resource Sharing**

Another important trend in the evolving relationship between universities and companies is the increased sharing of critical industry resources, including cloud computing, data, and open-source software. With the increasing emphasis on understanding how AI and deep learning technology can be applied to many domains, both inside and outside of computer science, key questions about what resources are needed to do high quality research arise. Modern industrial deep learning models, like the BERT language model recently published by Google[8], have hundreds of millions of parameters and require hundreds of days to train on the most powerful GPUs available.[9] Renting GPUs to compute such models using the AWS cloud costs literally tens of thousands of dollars for a single model training. Few, if any, faculty have access to computing at this scale without aligning with a company.

With AI deep learning research being both data and compute hungry, it becomes increasingly challenging for CS faculty to do cutting-edge research without partnering with companies. Such incentives will lead faculty to seek out collaborations and joint relationships with companies. At the same time, companies benefit from making dataset and cloud computing resources available to academics because encouraging academic research can drive marketplace competitiveness, especially when innovative research using the shared data aligns with the company's business goals. Ideally, companies and universities find a good balance where data is shared in pre-competitive scenarios in which companies can't justify the risk of productization and professors have the creativity and are willing to take risks to explore entirely new capabilities via their research.

Care must also be taken for intellectual property issues, especially in terms of informing students on if their research with a faculty member would be encumbered from an IP standpoint. Many approaches to handling IP are being utilized including automatic non-exclusive royalty free (NERF) licenses to a company from work done at the university by joint faculty member, faculty compartmentalizing their research (e.g., carve outs of what they are doing for the company), and keeping everything in the public domain. It is also important to anticipate situations where more than one university interacts with a company and IP and data sharing arrangements are needed between the three entities.

**Potential Opportunities and Guidelines**

We have identified an important trend in the evolution of the relation between the academic computing research community and industry and considered the positive and negative implications of this trend. We offer some suggestions and guidelines to spur further conversation on the topic and as a starting point to develop a shared approach in the computing research community.

---

[8] https://arxiv.org/abs/1810.04805
[9] http://timdettmers.com/2018/10/17/tpus-vs-gpus-for-transformers-bert/

**Thinking beyond technical engagements**: Most of the conversation to date has centered around the importance of industry engagement is area-specific (AI, IoT, health) and focused on access to technical subject matter knowledge, resources, and data. One of the undertones we heard in our interviews is the new lens academia can bring to industry from people that are at the intersection of these two worlds. We should think about how faculty can bring a societal lens that can have further impact. The reach of the principles and values from the academy informing products and R&D roadmaps has the potential to affect millions of people.  Beyond specific areas of computing, there are opportunities to cultivate shared knowledge, examine and discuss shared values and develop skills that can be mutually beneficial in industry and academia. There is a potential to build new trust from consumers if there is a clear academic influence on product. There is an opportunity to cultivate critical thinking skills, engage in discussions of ethical frameworks and ethical approaches to problem solving, and drive true efficacy.

**Disclosures**: Knowing which hat the faculty is wearing when is important, but even more important are **ALL** the hats they wear. In computing, we don't have the general culture of disclosing all of our affiliations or conflicts as in other disciplines like in medicine. This is complicated by the fact that some industry agreements have disclosure and communication limitations, which is something we found in our interviews. Disclosure is important so students, colleagues, and the community can make their own assessment and apply the right lens when a jointly appointed faculty member is presenting work. Particularly important is disclosure to students as there are IP restrictions (first rights, requirement to open source, etc), which can limit the areas they can engage. This is something that is important when recruiting and working with students, so they are clear on the guard rails for a particular topic area. "Sunlight is said to be the best of disinfectants." --Louis D. Brandeis

**Co-location and collaboration models:**
Companies have already observed that the most effective collaborations with universities happen when their staff are co-located. Hence numerous industrial "lablets" have been created for this purpose, including an investment by Intel of 6 lablets in the early 2000s.[10]  More recently companies such as Google[11], NVIDIA[12], and Uber[13] have opened labs adjacent to universities to increase contact and flow of ideas. Co-located lablets allow professors and students to more easily move between their university and industrial research and build stronger connections between the universities and companies. These have huge benefits to faculty, students, the university, and their region. Companies may establish new research laboratories adjacent to campus often if a faculty member is willing to lead the effort for a period of time. In these arrangements, industry can enable a faculty member to build a research group of a scale that could not be achieved using federal funding and that far exceed exceeds what the university might assemble on its own. They may also provide access to data and/or to computational resources that would not be available in the university setting, benefitting faculty, postdocs, and students.

---

[10] https://en.wikipedia.org/wiki/Intel_Research_Lablets
[11] https://www.princeton.edu/news/2018/12/18/google-open-artificial-intelligence-lab-princeton-and-collaborate-university
[12] https://news.developer.nvidia.com/nvidia-opens-robotics-research-lab-in-seattle/
[13] https://www.thestar.com/news/canada/2017/05/08/uber-opening-toronto-research-hub-for-driverless-car-technology.html

Historically, NSF and other government agencies have funded Centers of Excellence to enhance the combined research and teaching mission of universities and to attract greater participation from local industries (e.g., NSF's Cybersecurity Center of Excellence - CCoE - program[14]). The broader impact of computing technologies, including machine learning and AI, create the opportunity to create new, geographically dispersed, Centers of Excellence that combine universities, local industries, and investments from companies.

Finally, another opportunity to enhance the effectiveness of technology transfer between companies and universities is to create shared experiences that bring them together. In conversations with Norm Whitaker (Distinguished Scientist, Microsoft Research), former program manager for the DARPA Urban Challenge Autonomous Vehicle program[15], he reflected that the creation of a competition where companies and universities partnered together allowed companies to explore the technology in a pre-competitive environment. Because they were allied with universities, the prospect of losing the competition was less threatening because the company alone was not responsible for success or failure. He also reflected that the social context of teams of individuals working toward a common goal in a shared experience created a cadre of peer researchers that evolved into the technical teams making autonomous vehicles a reality today.

**Faculty arrangement guidelines:**
This emerging "new normal" of faculty members with deep extended engagements with industry brings new complexities that we will have to navigate as a community, which includes the impact that it has on the university culture and on the education and mentoring missions. This also presents new challenges from a university leadership standpoint where policies for computing faculty may start to differ than the rest of the university. It is clear that universities and departments may need to adapt to stay abreast of this emerging trend to keep up with research impact. It is clearly too early to stipulate rigid policies, but there are general principles that may provide a starting point. There is certainly a need for flexible experimentation, from which policies can be eventually be derived and guidelines agreed upon within the computing community.

Concurrent employment agreements must be in place to clearly delineate the faculty responsibilities to the university and company. We found that often faculty codes are at odds with certain industry employment agreements, especially around disclosures. It is also important for departments and the faculty member to make it clear to students if any IP generated by working with a faculty member engaged with industry is also encumbered, especially when recruiting new students (e.g., does a company have first rights to what is generated in an academic lab).

---

[14] https://www.nsf.gov/funding/pgm_summ.jsp?pims_id=505159
[15] https://en.wikipedia.org/wiki/DARPA_Grand_Challenge

In our informal interviews, we discovered the unintended consequence of students devaluing research in the academic setting from their observations of faculty deeply engaging in industry. Faculty engaged in industry should provide a balanced perspective, especially if they intend to continue to engage with the university. Visibility in the right context is also important. Do the students mainly see them with their industry hat on or their university hat or both? This also relates to mentorship as faculty with industry engagements will inherently have limited time for mentorship, teaching, and service. Some commitments need to be made by faculty if they intend to continue to engage with the university and these guidelines should be applied uniformly throughout the department or unit. Students should see faculty as faculty and not just as a member of industry.

The length of extended leaves for concurrent engagements is also a new challenge. Many universities have a 2-3 year limits on extended leaves, but many of these industry engagements require much longer arrangements to really have an impact (e.g., starting a new lab or building an entire new area of excellence). For extended leaves, a mentorship plan, service plan, teaching plans, conflict of interest management plan, and a timeline for the leaves must be clearly communicated to the university for planning purposes, but clearly there is uncertainty in many of these arrangements. University and departments must work closely to develop a viable plan, but recognize flexibility is needed as this is still an evolving model. Some departments have started to utilize without tenure (WOT) positions for indefinite industry engagements or a permanent reduction in their faculty FTE. Promotion and tenure committees will also need to define how to recognize impact from industry engagement.

## Summary and Next Steps

We observe that a significant shift is taking place and **computing research faculty are becoming more deeply engaged in working directly with companies as part of their research**. This trend will continue into the foreseeable future and that universities and companies will need to adjust in how they think about faculty, graduate students, and undergraduates based on this new reality.

The shift is motivated by important new opportunities that deep collaboration between faculty, students, and companies create and **many lasting benefits that result from these collaborations** including both improvements to research and community benefits from greater interaction between the cultures.

**This shift has the potential for negative impacts** on the kinds of research done, the quality of the research, the culture of computer science departments, and the training of undergraduates and graduate students. Particular attention needs to be focused on issues related to department culture, potential conflict of interest, intellectual property, and ensuring that students continue to have sufficient faculty mentoring and contact to prepare them for their career.

While some large universities have the resources and experience managing complex relations between jointly-appointed faculty and the university, we anticipate that **many departments, including some with fewer resources, will be faced with managing these relations**.

Awareness of the issues as well as guidance, including some of the principles mentioned in this report, will be a valuable resource in such cases. The CRA may be well positioned to inform the broad spectrum of computing research institutions about these trends and best practices in anticipating them, as they did in 2001.[16] See also other CRA best practices memos,[17] and especially the influential one with perspectives on conference publication for computing research tenure cases.[18]

Based on the preliminary perspective of this report, we hope that the CRA or similar authoritative body **creates a working group** focusing on taking our results forward, perhaps along the following lines:
- **Establish baseline measurements** and an ongoing process to monitor this trend, including a broad survey to understand the extent of the departments and subject areas impacted.
- Engage a committee of department chairs to **establish best practices** on getting the greatest benefit from joint appointments and collaborative agreements while anticipating the negative impacts outlined in this document.
- **Create a forum for computing research department chairs** to understand these trends, seek guidance and support in handling them, and share knowledge and experiences with others in the same situation.
- **Communicate these results to university administrators** and students so that their decisions can better reflect the new realities related to computing research departments.

## Acknowledgements

We received outstanding input on this topic from numerous members of the computing research community as a result of our discussions. We specifically thank Carl Andersen (US Dept. of Transportation), Charles Fay (US Dept. of Transportation), Mary Fernandez (MentorNet), Michael Franklin (Univ. of Chicago), Greg Hager (Johns Hopkins Univ.), Brent Hailpern (IBM), Mark Hill (Univ. Wisconsin), Karl Iagnemma (NuTonomy), Shriram Krishnamurthi (Brown Univ.), Vijay Kumar (Univ. Pennsylvania), Ed Lazowska (Univ. Washington), Erran Li (Columbia Univ.), Brian Noble (Univ. Michigan), Chris Ramming (VMware), Suresh Venkatasubramanian (Univ. Utah), Norm Whitaker (Microsoft), Lauren Wilcox (Google), colleagues at the National Science Foundation, and the members of the Computing Community Consortium Council (https://cra.org/ccc/about/ccc-council-members/) for their thoughtful comments.

*This material is based upon work supported by the National Science Foundation under Grant No. 1734706. Any opinions, findings, and conclusions or recommendations expressed in this material are those of the authors and do not necessarily reflect the views of the National Science Foundation.*---

[16] Commercialization Oversight for Computer Research Departments, CRA, 11/2001, https://cra.org/resources/best-practice-memos/commercialization-oversight-for-computer-research-departments/.

[17] CRA Best Practice Memos, https://cra.org/resources/best-practice-memos/.

[18] Evaluating Computer Scientists and Engineers For Promotion and Tenure, CRA, 9/1999, https://cra.org/resources/best-practice-memos/evaluating-computer-scientists-and-engineers-for-promotion-and-tenure/.